\documentclass[graybox]{svmult}
\pdfoutput=1
\usepackage{longtable}
\usepackage{mathptmx}      
\usepackage{helvet}
\usepackage{courier}       
\usepackage{type1cm}                                  
\usepackage{makeidx}         
\usepackage{graphicx}                                    
\usepackage{multicol}       
\usepackage[bottom]{footmisc}
\makeindex           
\begin{document}

\title*{The emergence of fermions and  the $\mathbf E_{11}$  content$^*$}
\author{Fran\c cois Englert and Laurent Houart}
\institute{Fran\c cois Englert \at   Service de Physique Th\'eorique and The International Solvay Institutes\\
 Universit\'e Libre de
Bruxelles, Campus Plaine C.P.225, 1050
Bruxelles, 
Belgium\\ \email{fenglert@ulb.ac.be}
\and Laurent Houart \at Service de Physique Th\'eorique et Math\'ematique and The International Solvay Institutes\\
 Universit\'e Libre de
Bruxelles, Campus Plaine C.P.231, 1050
Bruxelles, 
Belgium\\ \email{lhouart@ulb.ac.be}\medskip\\ $*$ Presented by F. Englert at the conference held on the occasion of Claudio Bunster's 60th birthday.}

\maketitle

\vskip -1cm
\abstract{Claudio's warm and endearing personality adds to our admiration for his achievements in physics a sense of friendliness. His constant interest in fundamental questions motivated the following presentation of our attempt to understand the nature of fermions. This problem is an essential element of the quantum world and might be related to the quest for quantum gravity. We shall review how space-time fermions can emerge out of bosons in string theory and how this fact affects the extended Kac-Moody approach to the M-theory project.}

\section{Introduction}

Despite the impressive theoretical developments of superstring theory, the quantization of gravity remains elusive. The difficulties encountered in coping with the non-perturbative level may well hide non-technical issues. A crucial point is whether the assumed quantum theoretical framework can cope with the quantum nature of space-time, in particular when confronted to the existence of black hole and the cosmological horizons. In this essay, we inquire into the fundamental nature of fermions, which constitute an essential element of the quantum world. We shall review how in string theory space-time fermions can be constructed out of bosons and we shall discuss how this fact affects the extended Kac-Moody approach to the M-theory project for quantum gravity.

In Section 2 we unveil the fermionic subspaces of the bosonic closed strings compactified on sublattices of a $E_8\times \widetilde{SO}(16)$ weight lattice, where $\widetilde{SO}(16)$ is the universal covering group of $SO(16)$~\cite{eht02}. All modular invariant fermionic closed strings, supersymmetric or not, are obtained from the parent bosonic strings by a universal truncation performed on both left and right sectors, or on the right sector for the heterotic strings. Supersymmetry arises when the sublattice of the $\widetilde{SO}(16)$ weight lattice is taken to be the $E_8$ root lattice so that the bosonic gauge group in each sector is ${\cal G}= E_8\times E_8$. We found that not only the closed string spectra of the fermionic string, but also the charges, the chiralities and the tensions of all the fermionic D-branes are encoded in the bosonic strings~\cite{eht02}.  In addition, the universal truncation applied to the unique tadpole-free unoriented bosonic string with Chan-Paton group $SO(2^{13})$ yields all tadpole- and anomaly-free open unoriented fermionic strings~\cite{eht01,eht02}.

In Section 3 we review the attempt to formulate the M-theory project in terms of the very-extended Kac-Moody algebra $E_{11}\equiv E_8^{+++}$~\cite{West:2001as} (or the overextended $E_{10}\equiv E_8^{++}$~\cite{Julia:1980gr}). Along this line of thought, the inclusion of the bosonic string suggest the introduction of the algebra $D_{24}^{+++}$~\cite{Lambert:2001gk} (or $D_{24}^{++}$). However it is easily shown that the $D_{24}^{+++}$ fields cannot accommodate the degrees of freedom  needed to generate the fermionic subspaces of the bosonic string. More generally we argue that, without extending the $D_{24}^{+++}$ algebra, one cannot encode genuine bosonic string degrees of freedom and, similarly, that $E_{11}$ alone does not encode genuine superstring degrees of freedom.

\section{Fermions and supersymmetry from the bosonic string}

It is well-known that ten-dimensional fermionic strings can be analyzed in
terms  of  bosonic operators, a consequence of the boson-fermion equivalence
in two dimensions. The approach taken here is different. We wish to
show that the Hilbert space of all the
perturbative fermionic  closed strings, and of all their tadpole- and anomaly-free open descendants, are subspaces of the
26-dimensional closed bosonic string theory, and of its tadpole-free open descendant, compactified on suitable
16-dimensional manifolds \cite{cent,ens,eht01,eht02}. 

\subsection{The fermionic subspaces of the closed bosonic string}

To accommodate space-time fermions  in the  left and/or the right sector of the 26-dimensional closed bosonic string one must meet three requirements:
\begin {enumerate}
\item{A continuum of bosonic zero modes must be removed. This
can be achieved by compactifying  $d=24-s$ transverse dimensions  on a
$d$-dimensional torus. This leaves $s+2$ non-compact dimensions with
transverse group
$SO_{trans}(s)$.}
\item{Compactification must generate an internal group
$SO_{int} (s)$ admitting spinor representations\footnote{Throughout this paper
we shall denote by $SO(s)$ all the groups locally isomorphic
to the rotational group of order $s(s-1)/2$. When specifically referring to its universal covering group, we shall
use the notation $\widetilde{SO}(s)$. Also we shall keep the same notation for groups and their Lie algebra.}. This
can be achieved by toroidal compactification on the weight lattice of a
simply laced Lie group
${\cal G}$ of rank $d$ containing a subgroup $SO_{int} (s)$. The
latter is then mapped onto  $SO_{trans} (s)$  in such a way that the
diagonal algebra
$ SO_{diag}(s) ={\rm diag} [SO_{trans}(s) \times SO_{int}(s)]$ becomes
identified with a new transverse algebra. In this way, the spinor
representations of  $SO_{int} (s)$ describe  fermionic states because
a rotation in space induces a half-angle rotation on these states. 
This mechanism is  distinct from the two-dimensional world-sheet
equivalence of bosons and fermions. It is reminiscent of a similar
mechanism at work in monopole theory: there, the diagonal subalgebra of
space-time rotations and isospin rotations can generate space-time
fermions from a bosonic field condensate in spinor representations of
the isospin group~\cite{thooft, jack, gold}.}
\item{The consistency of the above procedure relies on the
possibility of extending the diagonal algebra $SO_{diag}(s)$ to the
new full Lorentz algebra
$SO_{diag}(s+1,1)$,  a highly non trivial constraint. To break the
original Lorentz group $SO(25,1)$ in favor of the new one,   a
truncation consistent with conformal invariance must be performed on
the physical spectrum of the bosonic string. Actually, 
the states described by twelve compactified bosonic
fields must be projected out, except for momentum zero-modes of unit length 
\cite{cent,ens}. The removal of twelve bosonic fields accounts for the
difference  between the bosonic and fermionic light cone gauge central
charges. Namely, in units where the central charge of a boson is one, this
difference counts 11 for the superghosts and 
$(1/2) . 2 $ for time-like and longitudinal Majorana fermions. The  zero-modes
of length $\ell=1$ kept in the  twelve truncated dimensions contribute a
constant $\ell^2/2$ to the mass\footnote{We choose units in which the string length squared $\alpha^\prime=1/2$.}. They account for the removal by truncation of
the oscillator zero-point energies in these  dimensions, namely for
$-(-1/24).12=+1/2 $.  Moreover, the need to generate an
internal group $SO_{int}(s)$ via toroidal compactification requires
$s/2$ compactified bosons which can account for $s$ transverse
Majorana fermions (we hereafter take $s$ to be even, in which case
$s/2$ is the rank of the internal group). Therefore, one must ensure
that the total number
$d=24-s$ of compactified dimensions is at least $12 + s/2$. In other
words,
\begin{equation} \label{dimension} s\leq 8\ ,
\end{equation} and the     highest available space-time dimension
accommodating fermions is therefore
$s+2=10$ \cite{cent, ens}.
Here, we restrict our discussion to the case $s+2=10$.}
\end{enumerate}

To realize this program we choose a compactification of the
closed string at an enhanced symmetry point with gauge group ${\cal
G}_L\times {\cal G}_R$  where  
${\cal G}_L={\cal G}_R={\cal G}$ (or ${\cal G}_R={\cal G}$ for the heterotic string) and ${\cal G}=E_8\times SO(16)$ (or $E_8\times \widetilde {SO}(16)/Z_2=E_8\times E_8)$. Recall
that in terms of the left and right compactified momenta, the mass spectrum is
\begin{eqnarray} 
\label{spectrum} \frac{ m_L^2}{ 8}&=& \frac{ {\mathbf p_L}^2}{2}
+ N_L -1\, , \nonumber \\ \frac{ m_R^2}{ 8}&=&
\frac{ {\mathbf p_R}^2}{2} + N_R -1 \, ,
\\ &\hbox{and}&\nonumber\\ m^2 = { m_L^2\over 2}+{ m_R^2\over 2}\quad
&;&\quad m_L^2=m_R^2\, . \end{eqnarray} In Eq.(\ref{spectrum})
$N_L$ and
$N_R$ are the oscillator numbers in 26-dimensions and the zero-modes
${\mathbf p_L}$,
${\mathbf  p_R}$  span a
$2d$-dimensional even self-dual Lorentzian lattice with negative (resp.
positive) signature for left (resp. right) momenta. This ensures modular
invariance of the closed string spectrum \cite{narain}. The massless vectors
$\alpha_{-1,R}^\mu ~\alpha_{-1,L}^i\vert 0_L,0_R\rangle$ and
$\alpha_{-1,L}^\mu ~\alpha_{-1,R}^i \vert 0_L,0_R\rangle$, where the indices
$\mu$ and
$i$  refer respectively to non-compact and compact dimensions, generate for generic toroidal
compactification a
local symmetry $[U_L(1)]^d\times [U_R(1)]^d$. But more massless vectors
arise when $
{\mathbf p_L}$ and ${\mathbf p_R}$ are roots
of simply laced groups ${\cal G}_L$ and ${\cal G}_R$  of rank $d$ (with root
length
$\sqrt{2}$). The gauge symmetry is then enlarged to ${\cal G}_L\times {\cal G}_R$. 

\subsubsection{The group ${\cal G}= E_8\times \widetilde{SO}(16)$}

The compactification lattices in both sectors  (or in the
right sector only for the heterotic strings) are taken to be sublattices of the
${\cal G} =E_8\times \widetilde{SO}(16) $ weight lattice. These sublattices must preserve modular
invariance, which means that the  left and right compactified momenta
$\mathbf p_L$,
$\mathbf p_R$ must span a
$2d$-dimensional even self-dual Lorentzian lattice. 
All closed fermionic strings follow then from the properties of the $\widetilde{SO}(16)$ weight lattice and the subsequent truncation. 

The weight lattice $\Lambda_{\widetilde{SO}(2n)}$ split into four sublattices.
\begin{equation}
\label{so}
\Lambda_{\widetilde{SO}(2n)} =
\left\{\begin{array}{ll} (o)_{2n}:\mathbf p_o+\mathbf p&
(s)_{2n}: \mathbf p_s+\mathbf p \\
(v)_{2n}:\mathbf p_v+\mathbf p
& (c)_{2n}: \mathbf p_c+\mathbf p\\
\\
\mathbf p_o=(\underbrace{0,0,0,...0}_{n} ) & 
\mathbf p_s= (
\underbrace{{1\over2},{1\over2},{1\over2},...{1\over2}}_{n} )\\
\mathbf p_v= ( \underbrace{1,0,0,...0}_{n})&  \mathbf p_c= (
\underbrace{{-1\over2},{1\over2},{1\over2},...{1\over2}}_{n}) 
\end{array}\right\}
\end{equation}
and the $E_8$ weight (and root) lattice is
\begin{equation}
\Lambda_{E_8} =
(o)_{16} + (s)_{16}\, .
\end{equation}
Here $\mathbf p$ is a vector of the root lattice $\Lambda_R$ of $SO(16)$.

The partition functions $o,v,s,c$ corresponding to the lattices $(o),(v),(s),(c)$ are
\begin{equation}
\label{partition}
P_{j_{2n}}=\sum_{\mathbf p\in
\Lambda_R;N^{(c)}  }\exp 
\left\{2\pi i\tau\left[{(\mathbf p+\mathbf p_j)^2\over2}
 + N^{(c)} -{n \over 24 }\right]\right\} \quad j=o,v,s,c\, .
\end{equation}
Here $N^{(c)}$ is the oscillator number in the compact dimensions. Note that the additive group of the four sublattices of the weight lattice of $SO(2n)$ is isomorphic to the center of the covering group $\widetilde{SO}(2n)$, that is $Z_4$ for $n$ odd and $Z_2\times Z_2$ for $n$ even.

\subsubsection{The modular invariant truncation}

These lattices combine to form four modular invariant partition functions which after the truncation generate the four non-heterotic fermionic strings in ten dimensions~\cite{eht02}. In what follows, we shall only write down the $SO(16)$ characters in the integrand of amplitudes: the $E_8$ characters in  $E_8\times \widetilde{SO}(16)$ will be entirely truncated and will play no role and we do not display the contribution of the eight light-cone gauge non-compact dimensions. For the four bosonic ancestors, we get
\begin{eqnarray}
\label{BOB} OB_b&=&\bar o_{16}\ o_{16}+ \bar v_{16}\ v_{16}+
\bar s_{16}\ s_{16}+\bar c_{16}\ c_{16}\, \\ \label{BOA} OA_b&=&\bar o_{16}\
o_{16}+ \bar v_{16}\ v_{16}+
\bar s_{16}\ c_{16}+\bar c_{16}\ s_{16}\, \\
\label{B2B} IIB_b&=&\bar o_{16}\ o_{16}+ \bar s_{16}\ o_{16}+
\bar o_{16}\ s_{16}+\bar s_{16}\ s_{16}\, \\ \label{B2A} IIA_b&=&\bar o_{16}\
o_{16}+ \bar c_{16}\ o_{16}+
\bar o_{16}\ s_{16}+\bar c_{16}\ s_{16}\, 
\end{eqnarray}
where the bar superscript labels the left sector partition functions.

The universal truncation from $E_8\times \widetilde{SO}(16)$ to $SO_{int} (8)$ is defined by decomposing $SO(16)$ into $SO^\prime (8) \times SO(8)_{int}$ and erasing the $E_8$ and $SO^\prime (8)$ lattices except, in accordance with item 3 of the above discussion, for unit vectors of  $SO^\prime (8) $. In this way the internal momenta  $\mathbf p[SO(8)]$ are related to $\mathbf p[{\cal G}]$ by
\begin{equation}
\label{universal}
\framebox{$\displaystyle \frac{\mathbf p^2[{\cal G}]}{2}=\frac{\mathbf p^2[SO(8)]}{2} + \frac{1}{2}$}
\end{equation}
The unit vectors are identified as follows.
The decomposition of an
$SO(16)$ lattice in terms of $SO^{\,\prime}(8) \times SO(8)$ lattices yields
\begin{eqnarray} (o)_{16} = [(o)_{8^\prime}\oplus (o)_{8}] &+&
[(v)_{8^\prime}\oplus (v)_{8}] \, ,\nonumber\\ (v)_{16} = [(v)_{8^\prime}\oplus
(o)_{8}] &+& [(o)_{8^\prime}\oplus (v)_{8}]\nonumber\, ,\\ (s)_{16} =
[(s)_{8^\prime}\oplus (s)_{8}]& +&[ (c)_{8^\prime}\oplus
(c)_{8}]\nonumber\, ,\\  (c)_{16} = [(s)_{8^\prime}\oplus (c)_{8}] &+&
[(c)_{8^\prime}\oplus (s)_{8}]\, . \label{decompose}\
\end{eqnarray} The vectors of norm one in $SO^{\, \prime}(8)$ are the
4-vectors  $\mathbf p_v^\prime$, $\mathbf p_s^\prime$ and  $\mathbf p_c^\prime$ defined in Eq.(\ref{so}). We choose one
vector $\mathbf p_v^\prime$ and one vector $\mathbf p_s^\prime$. (One might equivalently have chosen $\mathbf p_c^\prime$ instead of $\mathbf p_s^\prime$.) 
In this way we get from Eq.(\ref{decompose})
\begin{eqnarray}
\label{truncations} &&o_{16} \rightarrow v_8\, , \qquad v_{16}\rightarrow o_8
\,
,\\ &&s_{16} \rightarrow -s_8\, ,
\quad ~ c_{16}
\rightarrow -c_8 \, .
\end{eqnarray} 
 It follows from the closure of the Lorentz algebra that states
belonging to 
$v_8$ or $o_8$ are bosons while those belonging to the spinor partition
functions $s_8$ and
$c_8$ are space-time fermions. The shift of sign in the fermionic amplitudes, which is consistent with the decomposition of $s_{16}$ and $c_{16}$  into $SO^\prime (8) \times SO(8)_{int}$, is required by the spin-statistic theorem and is needed to preserve modular invariance in the truncation. 
 
 The four ten dimensional fermionic string partition functions are
\begin{eqnarray}
\label{OB} OB_b \rightarrow \bar o_8\ o_8+ \bar v_8\ v_8+
\bar s_8\ s_8+\bar c_8\ c_8 &\equiv& OB\\ \label{OA} OA_b
\rightarrow
\bar o_8\ o_8+ \bar v_8\ v_8+
\bar s_8\ c_8+\bar c_8\ s_8 &\equiv& OA\\ \label{2B} IIB_b \rightarrow
\bar v_8\ v_8 - \bar s_8\ v_8 -
\bar v_8\ s_8+\bar s_8\ s_8 &\equiv& IIB \\
\label{2A} IIA_b \rightarrow \bar v_8\ v_8 - \bar c_8\ v_8 -
\bar v_8\ s_8+\bar c_8\ s_8 &\equiv& IIA
\end{eqnarray}
 Note that the partition functions of supersymmetric strings $IIA$ and $IIB$ arise from $E_8\times E_8$ sublattices of the $E_8\times \widetilde{SO}(16)$ weight lattice.
 
  The same procedure can be used to obtain all the heterotic strings, supersymmetric or not, by selecting the modular partition functions which are truncated in the right channel only. It will later be extended to D-branes and open descendants.

\subsubsection{The configuration space torus geometry}

The four modular invariant theories can be formulated in terms of the actions
\begin{eqnarray} S={-1\over 2
\pi}\int d\sigma d\tau \left[\{g_{ab}\partial_\alpha X^a\partial^\alpha X^b +
b_{ab} \epsilon^{\alpha \beta} \partial_\alpha X^a\partial_\beta X^b\}\right.
\nonumber\\
\left. +\eta_{\mu\nu}\partial_\alpha X^\mu\partial^\alpha X^\nu \right]\, ,
\label{action}
\end{eqnarray} with $g_{ab}$ a constant metric and $b_{ab} $ a constant
antisymmetric tensor in the compact directions ($a,b=1,...,16$),
$\eta_{\mu\nu} = (-1;+1,...)$ for
$\mu, \nu =1,...,10$ and
$0\leq \sigma \leq \pi$. The fields $X^a$ are periodic  with period $2\pi$. In this
formalism the left and right momenta are given by 
\begin{eqnarray} \mathbf p_R &=& [{1\over2} m_b + n^a (b_{ab} + g_{ab})] \mathbf
e^b,
\ \
\nonumber \\ \mathbf p_L &=&  [{1\over2} m_b + n^a (b_{ab} - g_{ab})] \mathbf e^b,
\label{close}
\end{eqnarray} where $\{\mathbf e^a\}$ is the dual of the basis $\{\mathbf
e_a\}$ defining the configuration space torus\footnote{In the previous sections momenta compactification was defined in both left and right channels. Both compactifications are obtained in the action formalism from the compactification on the configuration torus and the (quantized) $b$-field.} 
\begin{equation}
\label{ptorus} {\bf x} \equiv {\bf x} +2 \pi n^a \mathbf e_a \qquad n^a \in {\cal Z}\ , 
\end{equation} and the   lattice metric is given by \begin{equation}
\label{pmetric} g_{ab}= \mathbf e_a. \mathbf e_b\ .
\end{equation} 
  Explicit forms of the
$g_{ab}$ and $b_{ab}$ tensors for the four models are given in \cite{eht02}.

The D9-branes pertaining to the four different bosonic theories compactified on  the $E_8 \times \widetilde{SO}(16)$  lattices provide an easy way to construct their configuration space tori. We shall find that these tori are  linked to each other through global
properties of the universal covering group $\widetilde{SO}(16)$.

The tree channel amplitudes ${\cal A}_{tree}$  
of  the D9-branes  are obtained from the torus
partition functions Eqs.(\ref{BOB})-(\ref{B2A}) by imposing  Dirichlet boundary
conditions on the compact space.  For open strings the latter   
 do not depend on
$b_{ab}$ and are given by
 \begin{equation}
\label{diro}
\partial_\tau X^a=0\, ,
\end{equation} where $\tau$ is the worldsheet time coordinate and
$\sigma$ the space one.  Using the worldsheet duality which interchanges the
r\^oles of
$\tau$ and
$\sigma$, these equations yield the following relation between the left  and
right momenta:
\begin{equation}
\label{dirc} \mathbf p_L- \mathbf p_R=0\, ,
\end{equation} as well as a match between left and right oscillators in the tree
channel.
 The conditions Eq.(\ref{dirc}) determine the closed strings which propagate in
the annulus amplitude. Imposing them on the four tori  amounts to keep {\em
all} characters which appear diagonally in   Eqs.(\ref{BOB})-(\ref{B2A}).  Up to a
normalization, the annulus amplitudes, written as  closed string tree
amplitudes, are
\begin{eqnarray}
\label{ATB} {\cal A}_{tree}(OB_b)&\sim &  (o_{16}+ v_{16}+ s_{16}+
c_{16})\, ,\nonumber\\ {\cal A}_{tree} (OA_b)&\sim&
(o_{16}+
 v_{16})\, ,\nonumber\\ {\cal A}_{tree}(IIB_b)&\sim&(o_{16}+
s_{16})\, ,\nonumber\\ {\cal A}_{tree}(IIA_b)&\sim&
~o_{16}\, . \end{eqnarray}
We express  suitably normalized ${\cal A}_{tree}$ as a loop amplitude
${\cal A}$ for a single open string (i.e. without Chan-Paton
multiplicity) by performing a change of variable and the S-transformation on the
modular parameter ($\tau \rightarrow -1/\tau$). The result is given in Table~1. 

\begin{table}
\caption{Characters of the bosonic D9-branes in the tree and loop channels.}
\begin{tabular}{p{2cm}p{4cm}p{3cm}}
\hline\noalign{\smallskip}
&${\cal A}_{tree}$ & ${\cal A}$ \\
\noalign{\smallskip}\svhline\noalign{\smallskip}
$OB_b$ & $\sim (o_{16}+ v_{16}+ s_{16}+ c_{16})$  & $o_{16}$ \\
$OA_b$ &  $\sim (o_{16}+ v_{16})$&$o_{16}+ v_{16}$\ \\
$IIB_b$ & $\sim (o_{16}+ s_{16})$&$o_{16}+ s_{16}$\\
$IIA_b$ & $\sim \, o_{16}$ & $o_{16}+ v_{16}+ s_{16}+ c_{16}$ \\
\noalign{\smallskip}\hline\noalign{\smallskip}
\end{tabular}
\end{table}

The configuration space tori
of the four bosonic theories Eqs.(\ref{BOB})-(\ref{B2A}) are defined by
lattices with basis   vectors
$\{ 2\pi \mathbf e_a\}$ according to Eq.(\ref{ptorus}). We note that the Dirichlet
condition Eq.(\ref{dirc}) reduces Eq.(\ref{close}) to $\mathbf p_L =\mathbf p_R =
(1/2) m_a \mathbf e^a$ (independent of $b_{ab}$). Using the general expression
for lattice partition functions Eq.(\ref{partition}), we then read off for each
model the dual of its $SO(16)$ weight sublattice from the four tree amplitudes in  Table~1.  
  We then  deduce the $\{\mathbf e_a\}$ from the duality between the root lattice 
$(o)_{16}$ and the weight lattice
$(o)_{16}+(v)_{16}+(c)_{16}+(s)_{16}$, and from the self-duality of
$(o)_{16}+ (v)_{16}$ and $(o)_{16}+(s)_{16}$. We get
\begin{equation}
\label{pbase} \mathbf e_a= (1/2)\mathbf w_a\, ,
\end{equation} where the $\mathbf w_a$ are weight vectors forming a basis of a
sublattice $(r)_{16}$ of the weight lattice of $SO(16)$. The sublattice
$(r)_{16}$ for each theory is \begin{eqnarray}
\label{sublattice} (OB_b)& :& (r)_{16}=(o)_{16}\, , \nonumber\\ (OA_b)& :& 
(r)_{16}=(o)_{16}+(v)_{16}\, ,\nonumber
\\ (IIB_b)& : &(r)_{16}=(o)_{16}+(s)_{16}\, ,\nonumber \\ (IIA_b)& :&
 (r)_{16}=(o)_{16}+(v)_{16}+(s)_{16}+(c)_{16}\, .
\end{eqnarray}
These tori can be visualized by the projection depicted in Fig~1.

\begin{figure}[h]
\sidecaption
\includegraphics[width= 7.5 cm]{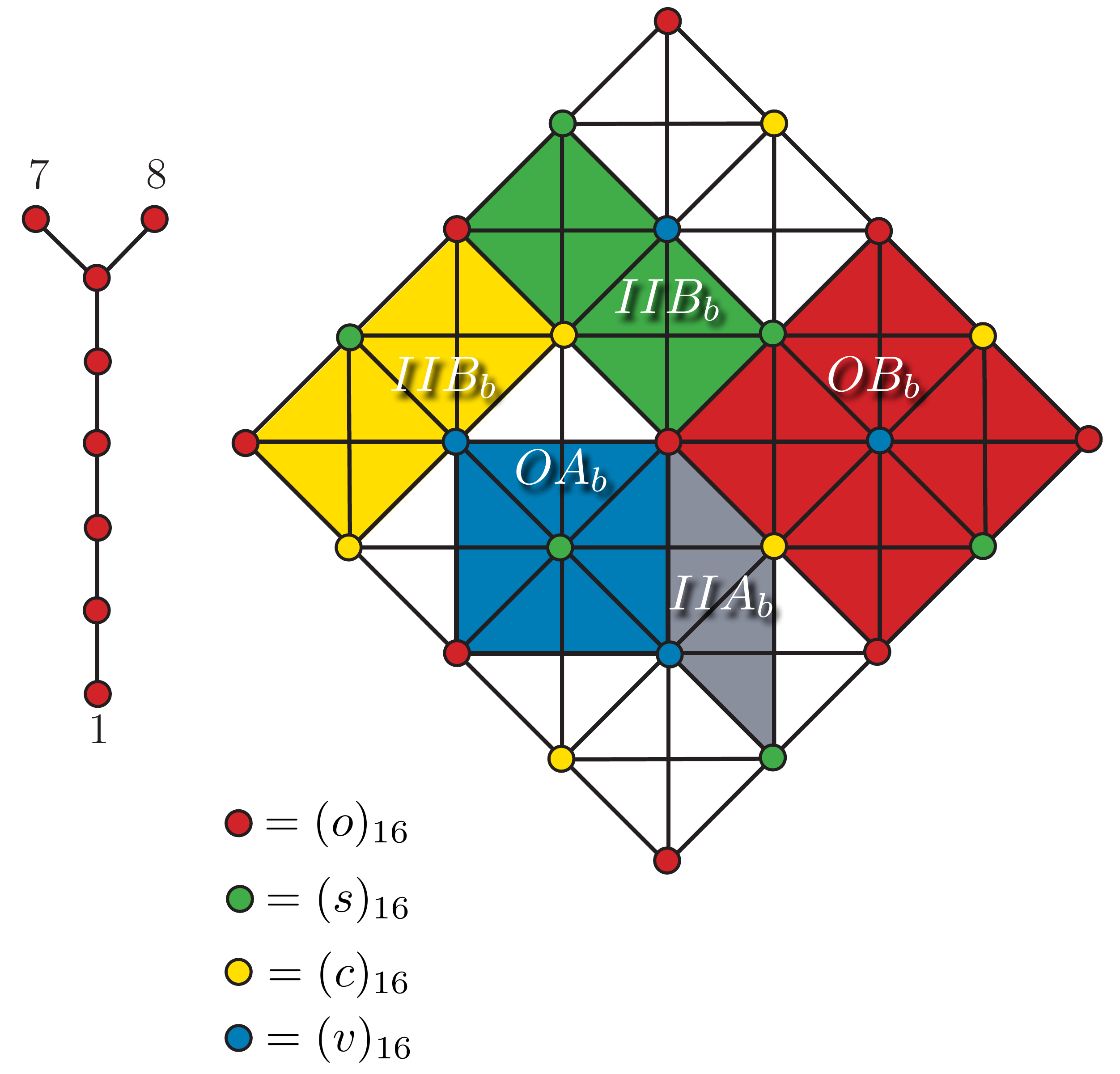}
\caption{Projected weight lattice of $\widetilde{SO}(16)$ in the
$7-8$ plane of the $SO(16)$ Dynkin diagram depicted in the figure.  We see from Eqs.(\ref{ptorus}), (\ref{pbase}) that the
volumes
$\xi_i$ of the unit cells, exhibited in shaded areas, must be multiplied by
$(2\pi)^8. 2^{-8}$ to yield the 
$SO(16)$ compactification space torus volume of the  four bosonic theories  (in
units where
$\alpha^\prime =1/2$). The two $IIB_b$ theories defined by the two rectangles are isomorphic and differ by
the interchange of $s_{16}$ and $c_{16}$.}
\label{fig:1}      
\end{figure}

The   tori $\widetilde t$ of the four bosonic theories are, as group spaces, the  
maximal toroids $\widetilde{\cal T}/Z_c $ of the locally isomorphic groups 
$E_8\times \widetilde{SO}(16)/Z_c$ where $Z_c$ is a subgroup of the centre 
$Z_2\times Z_2$ of the
 universal covering group
$\widetilde{SO}(16)$.  We write
\begin{eqnarray}
\label{coset}\widetilde t\, ( OB_b)& = &\widetilde{\cal T}\, ,\nonumber\\
\widetilde t\, (OA_b)& =&
\widetilde{\cal T}/Z^d_2\, , \nonumber\\ \widetilde t\, (IIB_b)& =
&\widetilde{\cal T}/Z^+_2 ~~~{\rm or}~~~
\widetilde{\cal T}/Z^-_2\, , \nonumber\\ \widetilde t\, (IIA_b)& =&
\widetilde{\cal T}/(Z_2 \times Z_2) \, , \end{eqnarray} where
$ Z^d_2 = diag(Z_2 \times Z_2)$ and the superscripts $\pm$ label the two
isomorphic $IIB_b$ theories obtained by interchanging $(s)_{16}$ and
$(c)_{16}$. 

There is thus a unified picture for the four theories related to the global
properties of the $SO(16)$ group~\cite{eht02}. The $OB_b$ theory built upon
$\widetilde{\cal T}$
 plays in some sense the role  of the `mother theory' of
the others.  One may view the different maximal toroids Eq.(\ref{coset}) as
resulting from the identification of  centre elements of 
$\widetilde{SO}(16)$, which are represented by   weight lattice points, with its
unit element. These identifications give rise  to the smaller shaded cells of
Fig.2.  In this way, the  unit cell of the
$IIB_b$ theory is obtained from the
$OB_b$ one by identifying the $(o)_{16}$ and $(s)_{16}$  lattice points (or
alternatively the 
$(o)_{16}$ and the
$(c)_{16}$ lattice points) and therefore
 also the $(v)_{16}$ and $(c)_{16}$ lattice (or
 the $(v)_{16}$ and the $(s)_{16}$ lattice),  as seen in Fig.2. It is therefore equal
to the unit cell of the
$E_8$ lattice\footnote{The latter however does not contain  $(v)_{16}$ and $(c)_{16}$
lattice points.}.   The  unit cell of the
$OA_b$ theory is obtained by identification of $(o)$ and $(v)$, and  of
$(s)_{16}$ and $(c)_{16}$.

\begin{figure}[h]
\includegraphics[width= 11.6 cm]{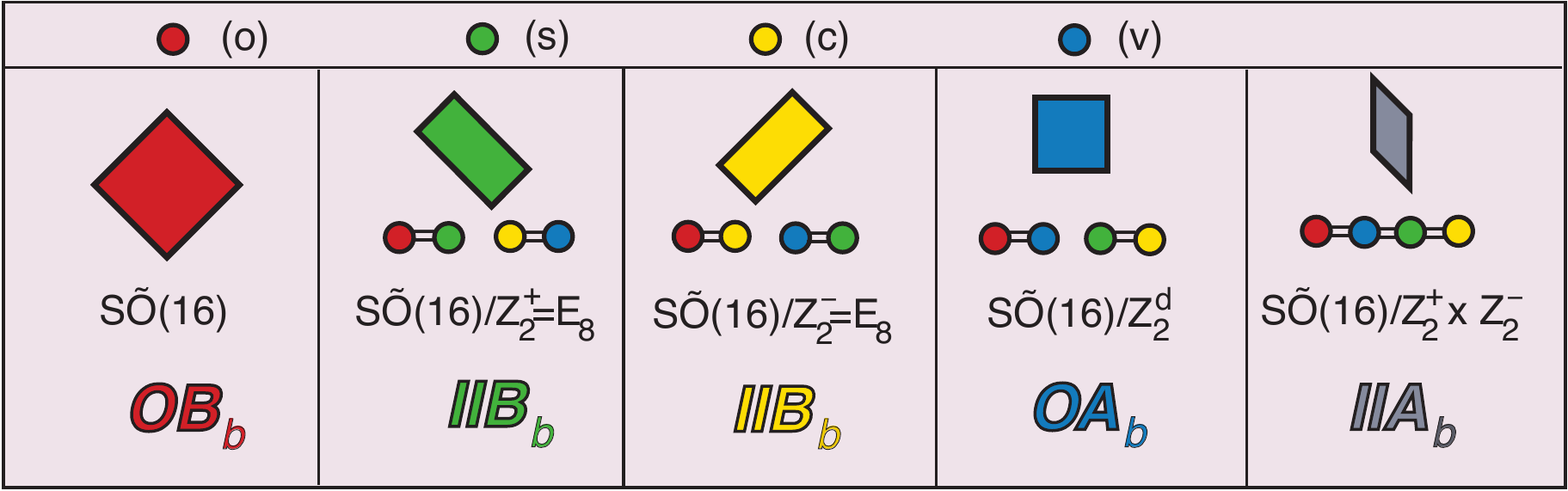}
\caption{Identification of  centre elements of 
$\widetilde{SO}(16)$ in the four closed string bosonic theories.}
\end{figure}

\subsection{The fermionic D-branes}

Table~1 lists the partition functions of a single  `elementary'  D9-brane for 
the four $SO(16)$ bosonic strings. In Table~2 below, using the universal truncation, we list the corresponding fermionic D9-branes and their loop partition function ${\cal A}_{trunc}$. We now generalize the analysis to encompass several D-branes~\cite{eht02}.

First, we remark that the relative position of the different
D9-branes in the eight compact dimensions of the $SO(16)$ torus is not arbitrary. Group symmetry requires that the partition function of
an open string with end points on D9-branes  be a linear combination of the
four $SO(16)$ characters. The vector {\bf d} separating the two points
where two distinct D9-branes meet the $SO(16)$ torus determines the
partition function of the  string starting at one point and ending at the other
after winding any number of times around the torus. The smallest eigenvalue
of the string Hamiltonian is $(1/2)\, {\bf d}.{\bf d}/\pi^2$. Therefore ${\bf d}/\pi$
must  be a weight and the D9-branes can only be separated in the compact
space (rescaled by a factor
$\pi^{-8}$) by a weight vector.  Consider for  instance two branes
 in the $OB_b$ theory, one located at $(o)_{16}$ and the other  located at $(v)_{16}$. The partition function of a string beginning and ending on the same brane is
$o_{16}$, while  the partition function of an open string stretching between
them  is $v_{16}$.  For the other theories, the partition function of a string
beginning and ending on the same brane will then contain, in addition to
$o_{16}$, the characters corresponding to the strings stretched between
$(o)_{16}$ and all points identified with $(o)_{16}$.  This can be checked by comparing the identifications indicated in Fig.2 with the partition functions $\cal A$ listed in
Table~1. 

If one chooses the location of  one elementary brane as the origin of the 
 weight lattice,  the other D9-branes can then only meet the $SO(16)$ torus
(rescaled by
$\pi^{-8}$) at a weight lattice point.  The number of distinct elementary branes
is, for each of the four bosonic theories, equal to the number of distinct
weight lattice points in the unit cell. For the mother theory
$OB_b$  there are four possible  elementary D9-branes. We label them by their
positions in the unit cell, namely by
$ (o)_{16}, (v)_{16}, (s)_{16}$  and $(c)_{16}$. Note that these weight lattice points  represent the
centre elements of the $\widetilde{SO}(16)$. 
 For the other theories the unit cells are smaller and there are fewer
possibilities. The unit cell of the
$IIB_b$ theory  allows  only  for  two distinct branes  $ (o)_{16}= (s)_{16}$,  $ (c)_{16}= (v)_{16}$ (or those obtained by the interchange of $(s)_{16}$ and $(c)_{16}$, as
seen in Fig.2).  Similarly for the $OA_b$ theory, we have the two branes
$(o)_{16}=(v)_{16}$ and
$(s)_{16}=(c)_{16}$, and finally for the `smallest' theory $IIA_b$, we have only one
elementary brane $(o)_{16}=(v)_{16}=(s)_{16}=(c)_{16}$. Finally to describe several D9-branes meeting at the same point of the $SO(16)$ torus, one uses the appropriate Chan-Paton factors. 

\subsubsection{Charge conjugation}

Charge conjugation of the truncated fermionic
strings is encoded in their bosonic parents. A brane sitting at
$(v)_{16}$ can always be joined by an open string to a brane sitting at $(o)_{16}$. The
partition function of such a string is given by the character $v_{16}$ and
therefore the two branes can exchange closed strings with tree amplitude 
${\cal A}_{tree} = o_{16}+ v_{16}-s_{16}-c_{16} $ as follows from the
S-transformation of the characters.  Namely the closed string exchange  
describing the interaction between these two
 branes has opposite sign for the $(s)_{16}$ and $(c)_{16}$ contribution  as compared to
 the closed string exchange between D9-branes located at the same point.
This shift of sign persists  in the truncation to the
fermionic theories where the  above tree amplitude becomes
$o_8+ v_8+s_8+c_8 $. This shift of sign   thus describes the  RR-charge conjugation between fermionic D9-branes.  It is encoded in the bosonic string as a  shift by the
lattice vector $(v)$  (see Fig.3). In particular, when $(o)_{16}$ and $(v)_{16}$ are
identified, all branes of the fermionic  offsprings are neutral. These are always
unstable branes, as the truncation of $v_{16}$ is
$o_8$ and contains a tachyon. Charged branes are always stable.

\begin{figure}[h]
\sidecaption
\includegraphics[width= 7.5 cm]{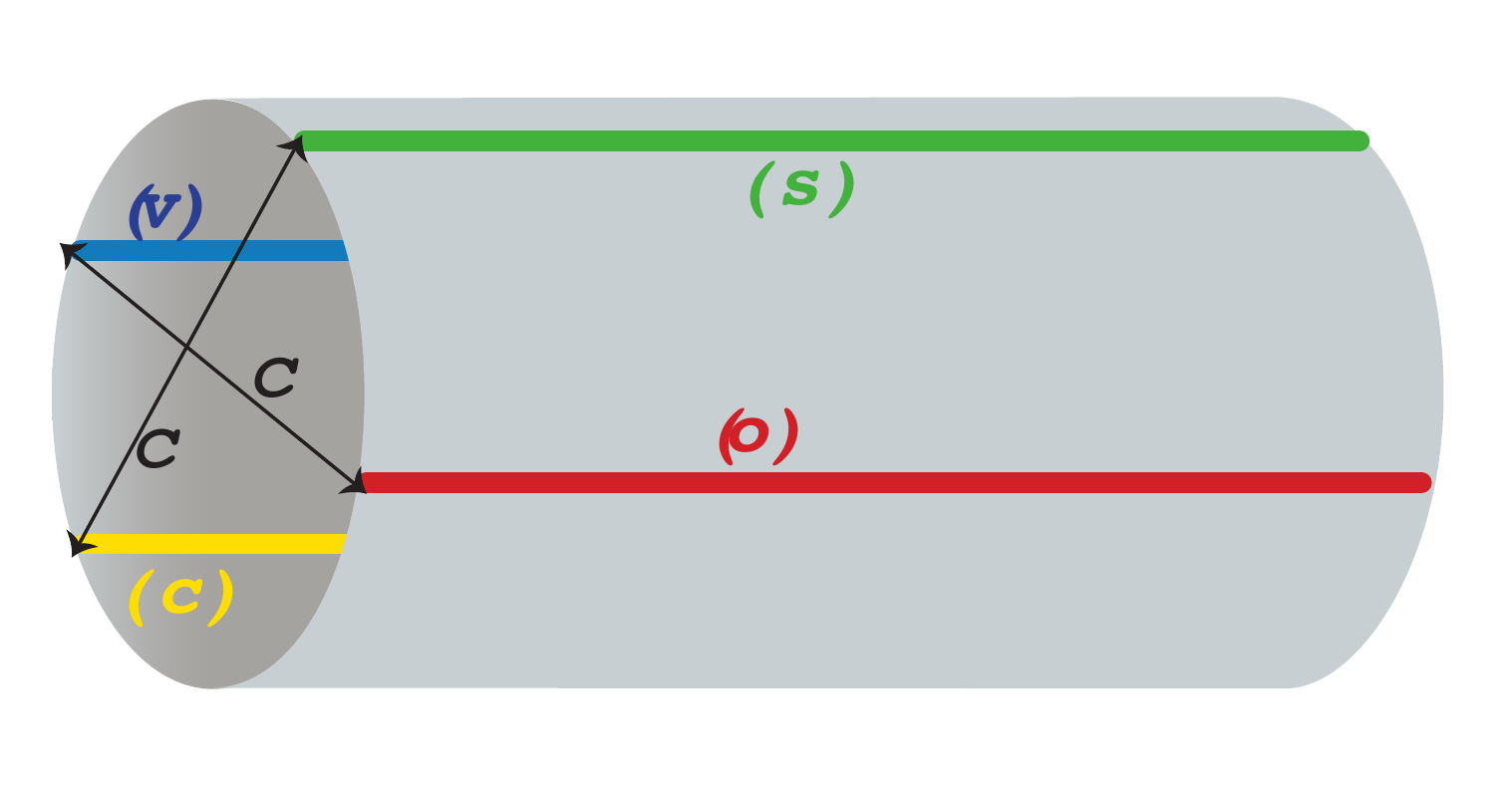}
\caption{{Charge conjugation of the fermionic D9-branes from the position of their bosonic ancestors on the $\widetilde {SO}(16)$ torus} (Subscripts in the labeling of the lattice points is omitted). The charge conjugate branes are linked by the lines $C$.}
 \end{figure}

The distinct fermionic D9-branes and their charge conjugates can
thus be directly read off from  Fig.1. They are included in Table~2,
where the charge is indicated by a superscript $+,-$ or $0$, and additional
quantum numbers by a subscript.

\begin{table}
\caption{Fermionic D9-branes charges encoded in their bosonic ancestors.}
\begin{tabular}{p{1.6cm}p{2.7cm}p{2.3cm}p{3.2cm}p{1.2cm}}
\hline\noalign{\smallskip}
&${\cal A}$ & ${\cal A}^{trunc}$&Fermionic D9-branes &Stability \\
\noalign{\smallskip}\svhline\noalign{\smallskip}
$OB_b\to OB$ & $o_{16}$  & $v_8$ &$ D_1^+ + D_2^+ +D_1^-  +D_2^-  $& stable\\
$OA_b\to OA$ &  $o_{16}+ v_{16}$&$v_8+ o_8$&$ D_1^0 + D_2^0$&unstable \\
$IIB_b\to IIB$ & $o_{16}+ s_{16}$&$v_8-s_8$&$D^+ +D^- $&stable\\
$IIA_b\to IIA$ & $o_{16}+ v_{16}+ s_{16}+ c_{16}$ & $v_8+ o_8-s_8- c_8$ &$D^0$&unstable\\
\noalign{\smallskip}\hline\noalign{\smallskip}
\end{tabular}
\end{table}

\subsubsection{Chirality}

We now consider the truncation of bosonic D-branes to lower 
dimensional fermionic Dp-branes (p$<$9). This is a non-trivial problem for  the
following reason. In fermionic string theories, a T-duality  interchanges type
$IIA$ with $IIB$, and type $OA$ with $OB$ while transmuting  D9-branes to 
D8-branes without changing their corresponding ${\cal A}^{trunc}$
amplitudes given in Table~2.  More generally, while the amplitudes of Dp-branes for $p$ odd are essentially the same as the D9-brane amplitude, those of the $p$ even would require, at the bosonic level, the interchange of $IIA_b$ with $IIB_b$ and $OA_b$ with $OB_b$ for the universal truncation to yield the correct chirality. 

This problem is beautifully solved~\cite{eht02} by noticing that the truncation of bosonic Dp-branes with $p$ even would violate Lorentz invariance for the fermionic strings except if this $A,B$ interchange could be performed at the bosonic parent level. And such interchange is indeed a symmetry of the compactified bosonic string! One has to perform an ``odd" E-duality which interchanges on the $SO(16)$ torus Dirichlet with Neumann boundary conditions and {\em simultaneously} performs the required $A,B$ interchange. One thus obtains in this way the loop amplitudes
of the bosonic parents which lead  from the universal truncation to the correct fermionic amplitudes consistent with Lorentz invariance. This is summarized in Table 3.

\begin{table}
\caption{Fermionic Dp-brane partition functions ($p\le 9$) and their bosonic ancestors.}
\begin{tabular}{p{1.5cm}p{2.5cm}p{2.5cm}p{2.2cm}p{2.2cm}}
\hline\noalign{\smallskip}
&${\cal A}_p,  p $ odd & ${\cal A}_{p+8},  p $ even &${\cal A}^{trunc}_p,  p $ odd&${\cal A}^{trunc}_p, p $ even \\
\noalign{\smallskip}\svhline\noalign{\smallskip}
$OB_b\to OB$ &  $o_{16}$  & $o_{16}+ v_{16}$&$v_{8}$&$o_{8}+v_{8}$\\
$OA_b\to OA$ &  $o_{16}+ v_{16}$& $o_{16}$ &$o_{8}+v_{8}$&$v_{8}$ \\
$IIB_b\to IIB$ &$o_{16}+ s_{16}$& $o_{16}+ v_{16}+ s_{16}+ c_{16}$&$v_{8}-s_{8}$&$o_{8}+v_{8}- s_{8}-c_{8}$\\
$IIA_b\to IIA$ & $o_{16}+ v_{16}+ s_{16}+ c_{16}$& $o_{16}+ s_{16}$ &$o_{8}+v_{8}- s_{8}-c_{8}$&$v_{8}-s_{8}$\\
\noalign{\smallskip}\hline\noalign{\smallskip}
\end{tabular}
\end{table}

\subsubsection{Tensions} 
We recall that the tension 
$T^{bosonic}_{Dp}$ of a Dp-brane in the 26-dimensional uncompactified theory
is  \cite{pobb}
\begin{equation}
\label{dpb} T^{bosonic}_{Dp} = {\sqrt\pi\over2^4 
\kappa_{26}}(2\pi\alpha^\prime{}^{1/2})^{11-p}\, , \end{equation}
 where
$\kappa_{26}^2= 8\pi G_{26}$ and $G_{26}$ is the Newtonian constant in 26
dimensions. The tensions of the Dirichlet D9-branes of the four compactified
theories are obtained from Eq.(\ref{dpb}) by expressing
$\kappa_{26}$ in term of the 10-dimensional coupling constant
$\kappa_{10}$. Recalling that $\kappa_{26}=
\sqrt{V} \kappa_{10}$ where $V$ is the volume of the configuration space
torus, one finds from Fig.1,
\begin{eqnarray}
\label{tob} T^{}_{OB_b} = {\sqrt\pi\over \sqrt{2}
\kappa_{10}}(2\pi\alpha^\prime{}^{1/2})^{-6}\, ,\\ \label{toa} T^{}_{OA_b} =
T^{}_{IIB_b} ={\sqrt\pi\over \
\kappa_{10}}(2\pi\alpha^\prime{}^{1/2})^{-6}\, ,\\ \label{t2a}
T^{}_{IIA_b}={\sqrt{2}
\sqrt\pi\over
\kappa_{10}}(2\pi\alpha^\prime{}^{1/2})^{-6}\, . 
\end{eqnarray} 
We now perform  the universal truncation  on  the loop
amplitudes ${\cal A}$ listed in Table~3 to compute the tensions of the fermionic branes.  Tensions are conserved in the truncation, as proven in
reference~\cite{eht01}. The tensions of the different bosonic D9-branes given in
Eqs.(\ref{tob})-(\ref{t2a}) are thus equal, when measured with the same
gravitational constant $\kappa_{10}$, to the  tensions of the corresponding
fermionic D9-branes \cite{pob,kets}. This is indeed a correct prediction. 

\subsection{Tadpole-free and anomaly-free fermionic open strings}

The open descendants of the closed bosonic
theories will be determined by imposing the {\it tadpole condition}
 \cite{sagr} on the bosonic string, namely by
imposing that divergences due to massless tadpoles cancel in the 
vacuum amplitudes. We will show that the bosonic $OB_b$,
$IIB_b$ and $OA_b$ theories admit tadpole-free open bosonic descendants and that those descendants give after truncation the three open fermionic string
theories which are anomaly- or tadpole-free~\cite{eht01,eht02}. Compactification of the bosonic string plays of course a crucial role, as the following analysis for the uncompactified unoriented bosonic string would recover the unique consistent Chan-Paton group $SO(2^{13})$.

\begin{figure}[h]
\includegraphics[width=8 cm]{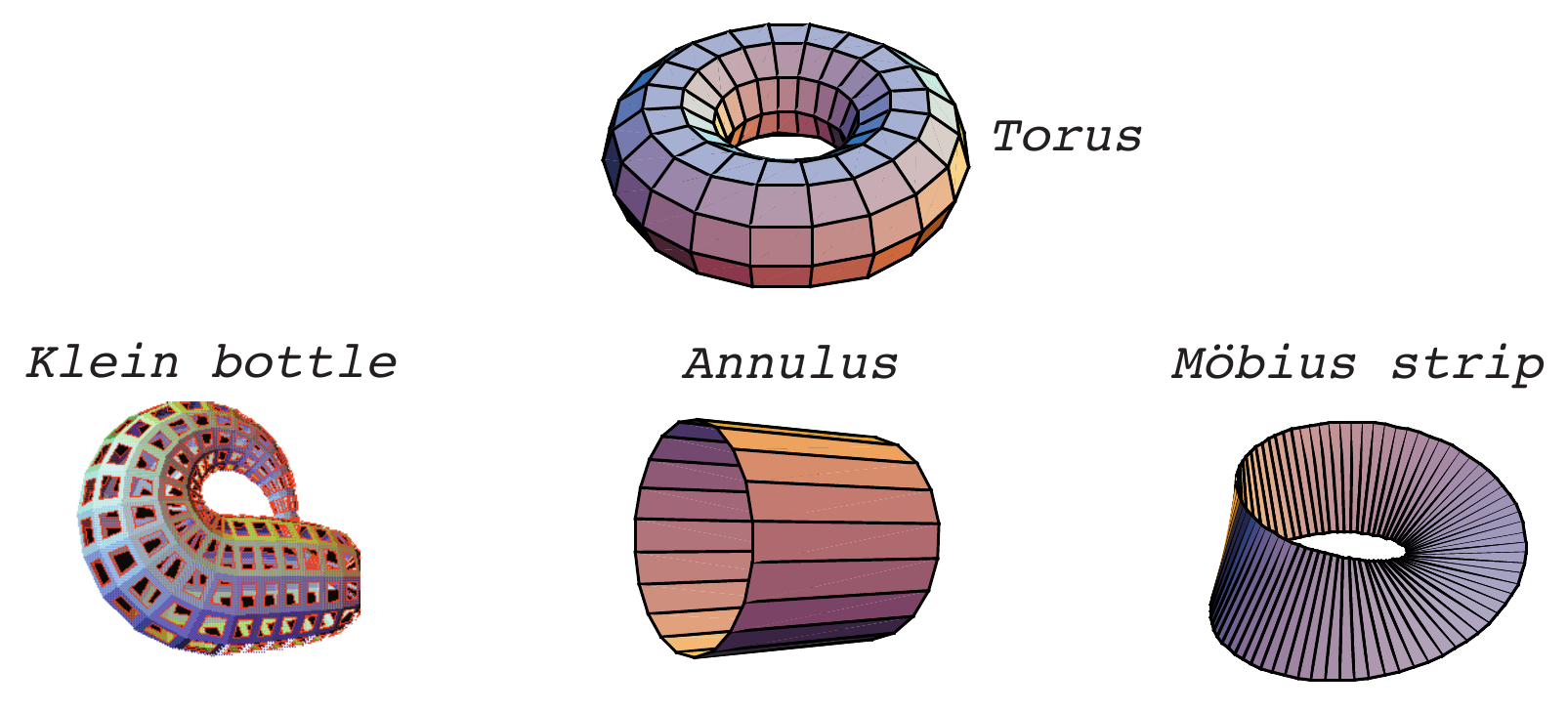}
\caption{World-sheet contributions to unoriented open strings.}
\end{figure}

A first step in obtaining the open descendant corresponding to the four
bosonic string theories characterized by the tori amplitudes ${\cal T}$
Eqs.(\ref{BOB})-(\ref{B2A}) is the construction of the  Klein bottle amplitudes
${\cal K}$. These are obtained from the amplitudes
${\cal T}/2+ {\cal K}$, which are the torus closed string partition functions
${\cal T}$ with the projection operator
$(1+\Omega)/2$ inserted, where $\Omega$ interchanges the left and right
sectors: $\Omega |L,R> =|R,L>$.  This can be done for $OB_b$, $IIB_b$ and $OA_b$ but not
for
$IIA_b$, because $\Omega$ in that case is not a symmetry of the theory. The
$IIA_b$ theory does not admit any open descendant. The
projection on  $\Omega$ eigenstates amounts  to impose the
 condition 
\begin{equation}
\label{kcon1} {\bf p_R = p_L}\, ,
\end{equation}   on the closed string momenta Eqs.(\ref{close}). 
Acting with
$\Omega /2$ on the three different tori Eqs.(\ref{BOB})-Eqs.(\ref{B2B}), one
finds the three Klein bottle amplitudes\footnote{Recall that we display only 
the
$SO(16)$ contribution  to the amplitudes.}
\begin{eqnarray}
\label{KOB} {\cal K}(OB_b) & = & {1 \over 2} (o_{16}+v_{16}+s_{16}+c_{16})\, ,\\
\label{K2B} {\cal K}(IIB_b) & = & {1 \over 2} (o_{16}+s_{16})\, ,\\ \label{KOA}
{\cal K}(OA_b) & = & {1 \over 2} (o_{16}+v_{16})\, .
\end{eqnarray}
The two remaining amplitudes with vanishing Euler characteristic,   the annulus
${\cal A}$ and the M\"obius strip
${\cal M}$, determine the open string partition function.  The
 annulus amplitudes of  D9-branes with generic Chan-Paton
multiplicities are 
\begin{eqnarray}
\label{cuob} {\cal A} (OB_b)&=& {1\over 2} (n_o^2 +n_v^2 + n_s^2 +n_c^2)
\, o_{16} + ( n_o n_v + n_s n_c) \, v_{16} \nonumber\\ & + & ( n_o n_s+ n_v n_c)
\, s_{16} +(n_o n_c + n_v n_s) c_{16}\, , \\
\label{cuoa} {\cal A}(OA_b)&=& {1\over 2} (n_o^2 + n_s^2) \, (o_{16} +
v_{16})  + n_o n_s \, (s_{16}+ c_{16})\, , \\
\label{cu2b} {\cal A} (IIB_b)&=& {1\over 2} (n_o^2 + n_v^2) \, (o_{16} +
s_{16}) + n_o n_v \, (v_{16}+ c_{16})\, .
\end{eqnarray}
To get the M\"obius amplitudes  ${\cal M}$ and to implement the tadpole
condition we express the Klein bottle  and annulus amplitudes
Eqs.(\ref{KOB})-(\ref{KOA}) and  Eqs.(\ref{cuob})-(\ref{cu2b}) as
closed string tree  channel amplitudes using the S-transformation of the
characters. From the resulting amplitudes  ${\cal K}_{tree}$ and  ${\cal A}_{tree}$ one obtains the  M\"obius amplitudes ${\cal M}_{tree}$ by requiring that each term in the power series
expansion of the total tree channel amplitude ${\cal K}_{tree}+{\cal A}_{tree}+
{\cal M}_{tree}$  be a perfect square. One gets 
\begin{eqnarray}
\label{MOB} {\cal M}_{tree}(OB_b) & = & \epsilon_1 \, (n_o +n_v + n_s +n_c)
\, {\hat o}_{16}\, ,\\
\label{M2B} {\cal M}_{tree}(IIB_b) & = & \epsilon_2 \, (n_o + n_v) \, {\hat
o}_{16} +
\epsilon_3 \, (n_o -n_v) \, {\hat s}_{16}\, , \\ \label{MOA} {\cal M}_{tree}(OA_b)
& = &
\epsilon_4 \, (n_o + n_s) \, {\hat o}_{16}  + \epsilon_5 \, (n_o- n_s) \, {\hat
v}_{16}\, , 
\end{eqnarray}  where $\epsilon_i = \pm 1$ will be determined by tadpole
conditions. The `hat' notation in the amplitudes Eqs.(\ref{MOB})-(\ref{MOA})
means that the overall phase present in the characters ${r}_{16}$ is
dropped. This phase arises because the  modulus   over which ${\cal M}$ is
integrated (and which is not displayed here) is not purely imaginary but  is
shifted by 1/2,  inducing  in  the partition functions $i_{16}$  an alternate shift
of sign in its power series expansion as well as a global phase. This one half shift is needed 
to preserve the group invariance of the amplitudes. A  detailed
discussion of the shift in general cases  can be found
in reference~\cite{sagr}.

We now impose the tadpole conditions on the three theories, namely we
impose the cancellation of the divergences due to the massless mode
exchanges in the total amplitudes
${\cal K}_{tree}+{\cal A}_{tree}+ {\cal M}_{tree}$. One determines in this way the Chan-Paton of the tadpole-free compactified unoriented bosonic open strings. The universal truncation preserves these factors and one gets in this way the correct tadpole-free and anomaly-free open fermionic strings~\cite{sagr}, as  indicated in Table 4. 

\begin{table}
\caption{Chan-Paton group of tadpole- and anomaly-free fermionic strings.}
\begin{tabular}{p{4cm}p{5cm}}
\hline\noalign{\smallskip}
&Chan-Paton group\\
\noalign{\smallskip}\svhline\noalign{\smallskip}
$OB_b\to OB\to B$ & $[SO(32-n)\times SO(n)]^2$ \\
$OA_b\to OA \to A$ &  $SO(32-n)\times SO(n)$\\
$IIB_b\to IIB\to I$ & $SO(32)$\\
$IIA_b\to IIA$ &--- \\
\noalign{\smallskip}\hline\noalign{\smallskip}
\end{tabular}
\end{table}

\subsection{The fermionic content of the bosonic string : summary}

\begin{svgraybox}
\begin{itemize}
\item[-]{From the torus compactification of the bosonic string on the weight lattice of $E_8\times \widetilde{SO}(16)$ and the universal truncation to $SO(8)$ keeping $\mathbf p_v^\prime,\mathbf p_s^\prime \in SO^\prime(8)$ one recovers all the closed fermionic string spectra. Those resulting from the compactification on the $E_8\times E_8$ sublattice are supersymmetric.}

\item[-]{Spectra, charges, chiralities, tensions of all the fermionic D-branes are obtained from their bosonic parents by the same universal truncation.}

\item[-]{Tadpole and anomaly cancellation of unoriented open fermionic strings follow from the tadpole free unoriented bosonic string by the same universal truncation.}
\end{itemize}
\end{svgraybox}

\section{The generalized Kac-Moody approach}

The five consistent superstring theories appear to be related by U-dualities and a conjectured non-perturbative formulation encompassing all of them has been labelled M-theory. Attempts to understand its symmetries has led to an approach to the M-theory project based on generalized Kac-Moody algebras. We shall analyze to what extend the connection between bosonic and fermionic strings found in Section 2 survives in this Kac-Moody approach. This will shed some light on its significance.

\subsection{$E_{11}\equiv E_8^{+++}$ and 11 dimensional supergravity}

Among the consistent superstring theories, type $IIA$ and type $IIB$ are maximally supersymmetric, i.e. they are characterized by 32 supercharges. We will focus on such maximally supersymmetric phase of  M-theory, whose classical limit is assumed to be  11-dimensional supergravity and whose dimensional reduction to ten dimensions yields the low energy effective action of type $IIA$ superstring.  The bosonic  action of 11-dimensional supergravity  is given by:
\begin{equation}
\label{Mth}
{\cal S}^{(11)} =\frac{1}{16\pi G_{11}}\,\int d^{11}x \sqrt{-g^{(11)}}\left(R^{(11)}- \frac{1}{2  . 4!}
F_{\mu\nu\sigma\tau}F^{\mu\nu\sigma\tau}+ CS \hbox  
{-term}\right).
\end{equation}
Scalars in the dimensional reduction of the action Eq.(\ref{Mth}) to three space-time dimensions realize non-linearly  the maximal non-compact form of the Lie group $E_8$  as a coset $E_8/SO(16)$ where $SO(16)$ is its maximal compact subgroup. Here, the symmetry of the (2+1) dimensionally reduced action has been enlarged from the $GL(8)$   deformation group of the compact torus $T^8$ to the simple Lie group $E_8$. This symmetry enhancement stems from the detailed structure of the action Eq.(\ref{Mth}). 

Coset symmetries were first found in  the dimensional reduction of 11-dimensional supergravity \cite{Cremmer:1978km}  to four space-time dimensions~\cite{Cremmer:1979up} but appeared also in other theories. They have been the subject of much study, and some classic examples are given in \cite{Ferrara:1976iq, Cremmer:1977tt, Julia:1980gr, Julia:1981wc, Cremmer:1978ds, Schwarz:1983wa}. In fact, all simple maximally non-compact Lie group $\cal G$   can be generated from the reduction down to three  dimensions from actions of gravity coupled to suitably chosen matter fields~\cite{Cremmer:1999du}. 

It has been suggested that these actions, or possibly some unknown extensions of them, possess a much larger symmetry than the one revealed by their dimensional reduction to three space-time dimensions in which all fields, except  $(2+1)$-dimensional gravity itself,  are scalars.  Such hidden symmetries would be, for each simple Lie group $\cal G$, the Lorentzian \cite{Gaberdiel:2002db} `overextended' $\cal G^{++}$ \cite{Damour:2002fz}  or `very-extended' $\cal G^{+++}$ \cite{Lambert:2001gk,Englert:2003zs} infinite Kac--Moody algebras generated respectively by  adding 2 or 3 nodes to the Dynkin diagram defining $\cal G$. One first adds the affine node, then a second node connected  to it by a single line to get the $\cal G^{++}$ Dynkin diagram and then similarly a third one connected to the second to generate $\cal G^{+++}$. In particular, the $E_8$ invariance of the
dimensional reduction to   three dimensions of 11-dimensional supergravity
would be enlarged to  $E_8^{++} \equiv E_{10}$ \cite{Julia:1980gr} or to
$E_8^{+++} \equiv E_{11}$, as first proposed in reference \cite{West:2001as}.   The extension of the Dynkin diagram of $E_8$ to $E_{11}$ is depicted in Fig.5. The horizontal line in Fig.5 form the Dynkin diagram of the $A_{10}$ subalgebra of $E_{11}$.  It is labelled {\it the gravity line},  as the nodes 4 to 10 of Fig.5 arise from the  reduction of gravitational part of the action Eq.(\ref{Mth}).  

\begin{figure}[h]
\sidecaption
\includegraphics[width=7.5 cm]{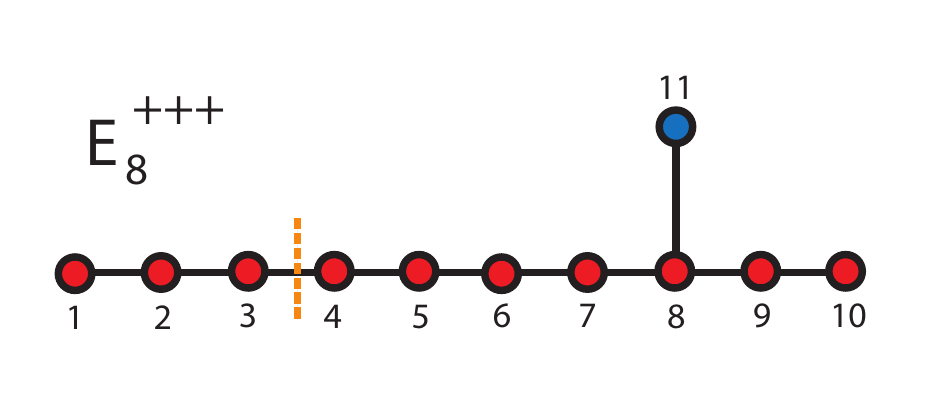}
\caption{Dynkin diagram of $E_{11}\equiv E_8^{+++}$.}
 \end{figure}   

To explore the possible significance of these huge symmetries a Lagrangian formulation \cite{Damour:2002cu}  {\it explicitly} invariant under $E_{10}$ has been proposed. It was constructed as a reparametrization invariant $\sigma$-model of fields depending on one parameter $t$, identified as a time parameter, living on the coset space $E_{10}/K_{10}^+$. Here $K_{10}^+$ is the subalgebra of $E_{10}$ invariant  under the Chevalley involution.  The $\sigma$-model contains an infinite number of fields and is built in a
recursive way  by a level expansion of $E_{10}$ with respect to its  subalgebra $A_9$ \cite{Damour:2002cu, Nicolai:2003fw}  whose Dynkin diagram is the `gravity line' defined in Fig.5, with the node 1  deleted\footnote{ Level expansions of $G^{+++} $ algebras in terms of a subalgebra $A_{D-1}$ have been considered in \cite{West:2002jj,Kleinschmidt:2003mf}.}.  The level of an irreducible representation of $A_9$ occurring in the decomposition of the adjoint  representation of $E_{10}$  counts the number of  times the simple root $\alpha_{11}$ not pertaining to the gravity line appears in
the decomposition.  The $\sigma$-model, limited to the roots up to level 3 and height 29, reveals  a perfect  match with the bosonic equations of motion of 11-dimensional supergravity in the vicinity of the  space-like singularity of the cosmological billiards \cite{Damour:2000hv,Damour:2001sa,Damour:2002et}, where fields depend only on time.  It was conjectured that   space derivatives are hidden  in some higher level fields of the $\sigma$-model~\cite{Damour:2002cu}. We shall label  this $\sigma$-model $S^{cosmo}$. 

An alternate $E_{10}$ $\sigma$-model parametrized by a space variable $x^1$ can be formulated on a coset space $E_{10}/K_{10}^-$, where $K_{10}^-$ is invariant under a `temporal' involution ensuring the Lorentz invariance $SO(1,9)$ at each  level in the $A_9$ decomposition of $E_{10}$.  This $\sigma$-model provides a natural framework for  studying static solutions~\cite{Englert:2003py, Englert:2003pz}. It yields all the basic BPS solutions of 11D supergravity, namely the KK-wave, the M2 brane, the M5 brane and the KK6-monopole, smeared in all space dimensions but one, as well as their exotic counterparts. We shall label the action of this $\sigma$-model $S^{brane}$.  The algebras $K_{10}^+$  and $K_{10}^-$  are both subalgebras of  the algebra $K_{11}^-$ invariant under the temporal involution defined on $E_{11}$, which selects the Lorentz group  $SO(1,10)=K_{11}^-\cap A_{10}$ in the $A_{10}$ decomposition of $E_{11}$~\cite{Englert:2003py,Englert:2004ph}. 

The underlying algebraic structure in this approach is thus $E_{11}$  and the infinite number of covariant fields are the parameters of the coset $E_{11}/K_{11}^-$ which can be recursively determined by the level decomposition with respect to $A_{10}$. In this section, we adopt this algebraic description of the field content of Kac-Moody algebras.

The first three levels contain the space-time degrees of freedom of 11-dimensional gravity along with their duals as depicted in Table 5. These levels are labelled classical. 

\begin{table}
\caption{Low level $E_{11}$ fields.}
\begin{tabular}{p{2cm}p{3cm}p{3.5cm}}
\hline\noalign{\smallskip} Level 
&field&supergravity content\\
\noalign{\smallskip}\svhline\noalign{\smallskip}
Level 0 & $g_{\mu\nu}$&gravity \\
Level 1&  $A_{\mu\nu\lambda}$&3-form potential\\
Level 2& $A_{\mu\nu\lambda\rho\sigma\tau} $& 6-form dual potential \\
Level 3&$ A_{\mu\nu\lambda\rho\sigma\tau\upsilon\zeta | \zeta} $&dual graviton\\
Level $\ge$ 3& non ``classical''&? \\
\noalign{\smallskip}\hline\noalign{\smallskip}
\end{tabular}
\end{table}

In this $E_{11}$ algebraic approach the crucial problem is to elucidate the role of the huge number of fields beyond the classical levels (level$\ge$  3, height $>$ 29)   and to find their significance.
Addressing this problem could bring an answer to two fundamental questions of this approach. First, is space-time itself encoded in the algebra and is  then  $E_{11}$ a symmetry of the uncompactified theory? Second, does  $E_{11}$ describes  the degrees of freedom of 11-dimensional maximal supergravity or more, for instance string degrees of freedom in some tensionless limit?  

It is fair to say that up to now there is no clear and totally satisfying answer to these questions. But recent progress to be discussed below points toward an answer to the second question. 

It has been  conjectured that the fields corresponding to the real roots of $E_9\subset E_{11}$ are dual fields and are not new degrees of freedom~\cite{Riccioni:2006az}. It was indeed shown that these fields express non-closing Hodge-like dualities relating between themselves the usual degrees of freedom of maximal 11-dimensional supergravity. Explicitly, from the $E_{11}$ fields parametrizing the coset $E_{11}/K^-_{11}$, the subset of real roots of $E_9\subset E_{11}$ generate,  using these non-closing dualities  realized as $E_9$ Weyl reflections,  an infinite U-duality $E_9$ multiplet of BPS static solutions of 11-dimensional supergravity~\cite{Englert:2007pz}.

In another development ~\cite{Riccioni:2007au, Bergshoeff:2007qi}, it was shown that another class of $E_{11}$ fields contain all those needed to describe all the maximal gauged supergravity in $D \leq 11$ dimensions. Namely  the  $D-1$ forms and the $D$-forms content, present in the $E_{11}$ algebraic description interpreted in $D$ dimensions, matches the embedding tensor description~\cite{deWit:2008ta} of all the  gauged maximal supergravities. Hence the $E_{11}$ algebraic approach appears to contain the algebraic structure of all maximal non-abelian supergravities (with 32 supercharges). However again, although these transcend 11D ungauged maximal supergravity they do not contain new degrees of freedom. 

There is still an infinite number of other fields characterized by $A_{10}$ representations with mixed Young tableaux. Their significance is hitherto unclear.

We now turn to the generalized Kac-Moody approach to the bosonic string. This will give some new information on the field content in the algebraic approach.

\subsection{$D_{24}^{+++}$ and the bosonic string}

The low-energy effective action of the
$D=26$ bosonic string (without tachyon) contains gravity,  the NS-NS three form field strength and the dilaton. It is given by
\begin{equation}
 \label{lad}
    S =  \int d^{26}x \, \sqrt{-g}\left(R-
\frac{1}{2}\partial_\mu\phi\partial^\mu\phi-{1\over 2 . 3!
}e^{-\frac{1}{3}\phi}H_{\mu\nu\sigma}H^{\mu\nu\sigma}\right)\, ,
\end{equation} and $H=db$.

Upon dimensional reduction to three space-time dimensions, one would have expected to have
a $GL(23) \times U(1)$ symmetry (the U(1) coming from the dilaton). Again, there is  an enhancement to a simple Lie algebra, namely $D_{24}$ in its split form.  The symmetry is non-linearly realized and the scalar lives in the coset $SO(24,24)/SO(24) \times SO(24)$. The corresponding Dynkin diagram of $D_{24}$  is the part of the diagram  Fig.6  on the right of the dashed line.

\begin{figure}[h]
\sidecaption
\includegraphics[width=5.5 cm]{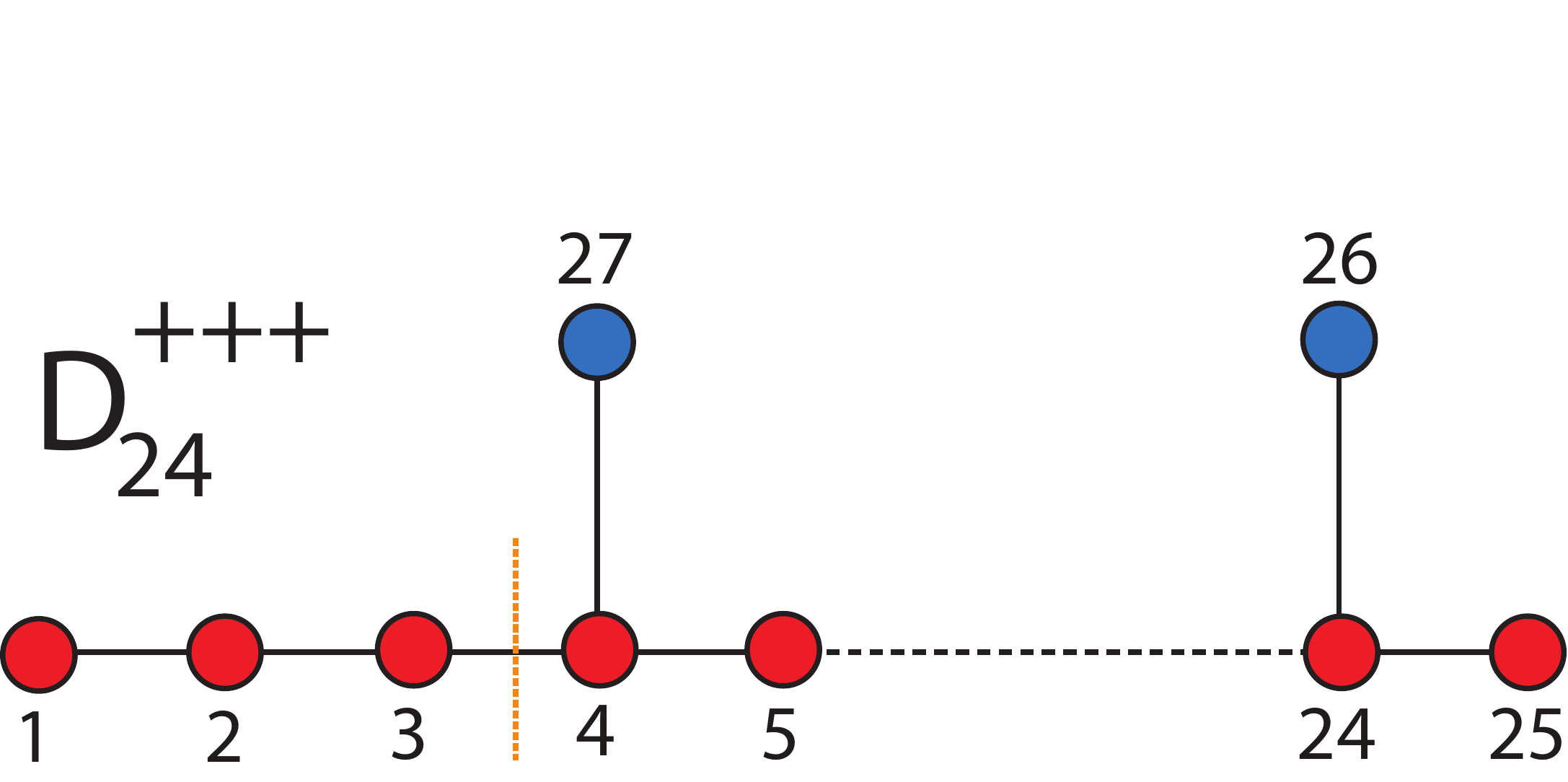}
\caption{Dynkin diagram of $D_{24}^{+++}$.}
 \end{figure}

Having this symmetry in 3 dimensions,  the discussion of the preceding section suggests that the `very extended' Kac-Moody algebra $D_{24}^{+++}$ could encode a symmetry  of the bosonic string~\cite{Lambert:2001gk}.  The physical fields of this algebraic approach would then live in a coset  $D_{24}^{+++}/K_{24}^-$ where $K_{24}^-$ is the maximal  subalgebra invariant under the temporal involution. The corresponding Dynkin diagram is depicted in Fig.6.

To make contact with the analysis of Section 2 and uncover a possible relation through truncation with the fermionic strings in 10 space-time dimensions we consider the decomposition of $D_{24}^{+++}$ into $A_9 \times D_{16}$ where the diagram of  $A_9$ is the gravity line of a  10 dimensional space-time and $D_{16}$ a symmetry arising from a torus compactification of 16 dimensions.  In this decomposition the unbroken subalgebra $K_{24}^-$ decomposes into $SO(1,9)  \times SO(16) \times SO(16)$. Physical fields appear in the double level decomposition with respect to the nodes $\alpha_{27}$ and $\alpha_{10}$ in Fig.7. 

\begin{figure}[h]
\includegraphics[width=10cm]{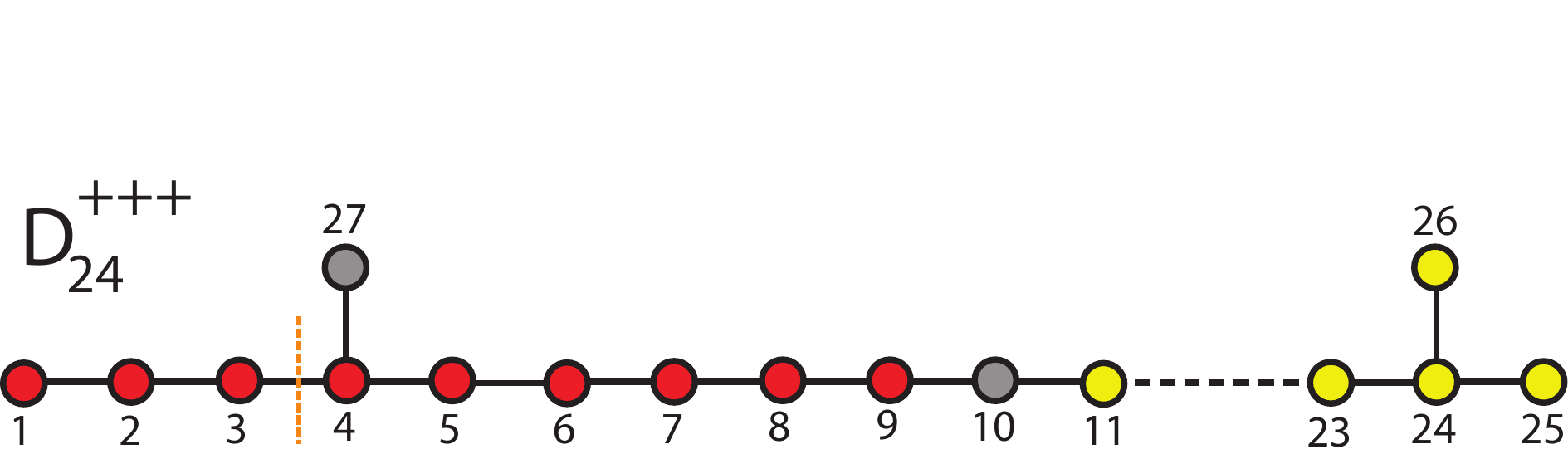}
\caption{ The decomposition of $D_{24}^{+++}$ into $A_9\times D_{16}$ with levels defined by the nodes $\alpha_{27}$ and $\alpha_{10}$.}
 \end{figure}

The level $l_1$ (resp. $l_2$) counts the number of times the root $\alpha_{10}$ (resp. $\alpha_{27}$) appears in the decomposition of the adjoint representation of $D_{24}^{+++}$ into  irreducible representations of $A_9$ and $SO(16,16)$. The first levels are listed in Table 6 obtained from the SimpLie program~\cite{nutma}.

\begin{table}
\caption{Low level representations in the decomposition of $D_{24}^{+++}$ into $A_9\times D_{16}$. Their Dynkin labels are $p_r$ and $p_i$, their dimensions $d_r$ and $d_i$. $r^2$ is the root norm and $\mu$ the outer multiplicity. }
\begin{tabular}{p{1cm}p{2cm}p{3cm}p{1cm}p{1.4cm}p{1.4cm}p{1cm}}
\hline\noalign{\smallskip}
$l_1, \,l_2$  & $p_i$&$p_r$ &$r^2$&$d_r$&$d_i$&$\mu$\\
\noalign{\smallskip}\svhline\noalign{\smallskip}
$0\, , \,0$&000000000& 000000000000000&0&1&1&2\\
$0\, , \,0$&100000001& 000000000000000&2&99&1&1\\
$0\, , \,0$&000000000& 010000000000000&2&1&496&1\\
\hline
$1\, , \,0$&000000001& 100000000000000&2&10&32&1\\
\hline
$0\, , \,2$&100000100& 000000000000000&2&1155&1&1\\
\hline
$2\, , \,0$&000000010& 000000000000000&2&45&1&1\\
\hline
$0\, , \,1$&000100000& 000000000000000&2&210&1&1\\
\hline
$1\, , \,1$&001000000& 100000000000000&2&120&32&1\\
\hline
\end{tabular}
\end{table}

Thus, the internal space of the
the physical fields is the coset $SO(16,16)/ SO(16) \times SO(16)$. This is exactly the moduli space of modular invariant compactifications of the closed bosonic string on a 16-dimensional torus. At a generic point,  one has $U(1)_L^{16} \times U(1)_R^{16}$ where L  (resp. R) stands for left (resp. right).  We thus expect to find 32 abelian gauge fields and these indeed appear in Table 6 at the level $(1,0)$ in the fundamental representation of $SO(16,16)$. As explained in Section 2, the compactifications needed for uncovering by truncation the fermionic strings are the special points of enhanced symmetry in the coset where the torus is identified to one of the four maximal toroids of $[\widetilde{SO}(16)/Z_i ]\times E_8$, where $Z_i$ is an element of the center $Z_2\times Z_2$ of the universal covering $\widetilde{SO}(16)$. For $Z_i = Z_2$, (resp. $ Z_i=Z_2\times Z_2$) this yields the gauge group $(E_8\times E_8)_L\times (E_8\times E_8)_R$, which yields after truncation to the maximal supersymmetric type IIB (resp. IIA) string theory.

One may first ask if the  non-abelian extension of the gauged  $U(1)_L^{16} \times U(1)_R^{16}$ to  $(E_8\times E_8)_L\times (E_8\times E_8)_R$, which appear at these enhanced symmetry points  are encoded in $D_{24}^{+++}$, as do the non-abelian gauging of maximal supergravities  in $E_8^{+++}$. The answer is no, as spinor representations of $D_{16}$ cannot appear in the adjoint representation of $D_{24}^{+++}$. This means that one would have to extend the algebra of $D_{24}^{+++}$ to include fields not contained in the adjoint representation of the generators if one wishes to recover the information encoded in the torus compactification at these enhanced symmetry points.

The problem is not limited to enhanced symmetry points involving spinor representations of the group. The stringy nature of the massless degrees of freedom enlarging the abelian gauging to a non abelian one at enhanced symmetry points has no counterpart in  the non-abelian gauging of maximal supergravities which appear in $E_8^{+++}$ and which are studied in reference~\cite{Riccioni:2007au,Bergshoeff:2007qi} . These indeed do not introduce new degrees of freedom. One might then expect that the $D_{24}^{+++}$ fields do not comprise the genuine string degrees of freedom of the bosonic string. Similarly, despite the fact that $E_8^{+++}$ does contain spinor representations of orthogonal groups, the $E_8^{+++}$ fields are not expected to comprise genuine superstring degrees of freedom such as the massless fields resulting from torus compactifications at enhanced symmetry points. In that case, if the the full set of string degrees of freedom are to be included in some M-theory project, its algebraic description would transcend the description by the $E_8^{+++}$ fields.

\begin{acknowledgement}

We are greatly indebted to  Axel Kleinschmidt and to Nassiba Tabti for fruitful discussions. One of us, Fran\c cois Englert, is grateful to Peter West for an interesting discussion on the possible extensions of $E_8^{+++}$.
We would  like to warmly thank  Marc Henneaux,  Vivian Scharager and Jorge Zanelli for having organized such friendly and magnificent ClaudioFest, which was made even more lively by the constant participation of Claudio himself. 

Laurent Houart is a Research Associate of the Fonds de la Recherche Scientifique-FNRS, Belgium. This work was supported in part by IISN-Belgium (convention 
4.4511.06 and convention 4.4505.86), and by 
the Belgian Federal Science Policy Office through the 
Interuniversity Attraction Pole P VI/11.

\end{acknowledgement}

%
%
%

\end{document}